\documentclass[aps,pra,reprint,amsmath,amssymb,amsfonts,longbibliography,superscriptaddress,floatfix]{revtex4-2}
\usepackage[english]{babel}
\usepackage[utf8x]{inputenc}
\usepackage[T1]{fontenc}
\usepackage[a4paper,top=3cm,bottom=2cm,margin = 25mm]{geometry}
\usepackage{bbold}
\usepackage{amsmath}
\usepackage{amssymb}
\usepackage{graphicx}
\usepackage{verbatim}
\usepackage{booktabs}
\usepackage{xcolor}
\usepackage{hyperref}
\newcommand{\ket}[1]{\left| #1 \right\rangle} 
\newcommand{\bra}[1]{\left\langle #1 \right|} 

\newcommand{\matel}[2]{
\ifnum 0=#1\relax
S_{#2}
\else
S_{#2}^{*}
\fi
}

\newcommand{\EdvinUpdate}[1]{\textcolor{black}{#1}}

\begin{document}
\title{Coherent control of ionization via stabilization by resonant pulse pairs}

\author{Edvin Olofsson}
\affiliation{Department of Physics, Lund University, Box 118, SE-221 00 Lund, Sweden}
\author{Evan Lovelle Fulton}
\affiliation{Department of Physics and Astronomy, University of Nebraska, Lincoln, Nebraska 68588-0299, USA}
\author{Rezvan Tahouri}
\affiliation{Department of Physics, Lund University, Box 118, SE-221 00 Lund, Sweden}
\author{Mattias Bertolino}
\affiliation{Department of Physics, Lund University, Box 118, SE-221 00 Lund, Sweden}
\affiliation{Institue of Theoretical Physics, Faculty of Mathematics and Physics, Charles University, V Hole{\v s}ovi{\v c}k{\' a}ch 2, Prague 8, 180 00 Czech Republic}
\author{Jean Marcel Ngoko Djiokap}
\affiliation{Department of Physics and Astronomy, University of Nebraska, Lincoln, Nebraska 68588-0299, USA}
\author{Jan Marcus Dahlström}
\email{marcus.dahlstrom@fysik.lu.se}
\affiliation{Department of Physics, Lund University, Box 118, SE-221 00 Lund, Sweden}


\begin{abstract}
    We study the nonlinear and resonant process of two-photon ionization of atoms (He and H) in a pump-probe scheme. The pump pulse prepares the quantum system in a superposition of the ground state and an excited bound state. By varying the phase difference between the pulses, we show how it is possible to coherently control the dressed-state population during the probe pulse. Our main result is that for certain laser parameters, the control over the dressed state population leads to strong control of the ionization probability during the probe pulse. The effect arises due to one of the dressed states becoming stabilized against ionization. 
    Contrasting effects from circular and linear polarized pulses demonstrate how such ``bound states in the continuum'' are sensitive to the degeneracy of the coupled continuum.
\end{abstract}

\maketitle


\section{Introduction}\label{sec:Introduction}

What happens to atoms in strong fields? 
At low frequencies, where more than one photon is required for photoionization, there are two separate theoretical approximations: the strong-field approximation (SFA) \cite{amini_symphony_2019} and the rotating-wave approximation (RWA) \cite{CT1998}. These approaches are complementary as they apply to non-resonant and resonant processes, respectively. In more extreme cases, atoms can be found in high-frequency superintense laser pulses, where one photon is enough for photoionization, but the forces from the laser field and the atomic potential are comparable \cite{gavrila_atomic_2002}. This implies that neither SFA nor RWA are applicable and that another route is required. In this case, quasiclassical approaches have been proposed, based on the Wentzel–Kramers–Brillouin (WKB) approximation \cite{fedorov_strong-field_1998,ivanov_semiclassical_1998}, and by performing perturbative Coulomb corrections to SFA \cite{popruzhenko_strong_2008}, but the simultaneous action of laser and atom can only be accounted for quantitatively by numerical propagation of the time-dependent Schr\"odinger equation (TDSE) \cite{eberly_atomic_1993,zimmermann_unified_2017} or Dirac equation (TDDE) \cite{kjellsson_relativistic_2017}. Interestingly, it has been found that all three strong-field regimes share the counterintuitive feature that ionization can be prevented (or reduced), a phenomenon commonly referred to as {\it stabilization}. 
Here, we focus on stabilization of atoms in strong fields, and, as proposed by Gavrila \cite{gavrila_atomic_2002}, we will distinguish between {\it quasistationary stabilization} (QS), which concerns decay rates obtained by Floquet-like theories for atoms in monochromatic fields, and {\it dynamical stabilization} (DS), which concerns time-dependent pulse shapes and atomic populations from TDSE (or TDDE) computations. \EdvinUpdate{In the QS sense, stabilization is characterized by a decrease or oscillatory behavior of ionization rates with increasing intensity \cite{gavrila_atomic_2002}. On the other hand, DS is characterized by the behavior of the ionization probability at the end of a pulse sequence, if the ionization probability levels out or decreases with an increase in the intensity \cite{gavrila_atomic_2002}.}

\begin{figure}[t]
    \centering
    \includegraphics[width=0.75\linewidth]{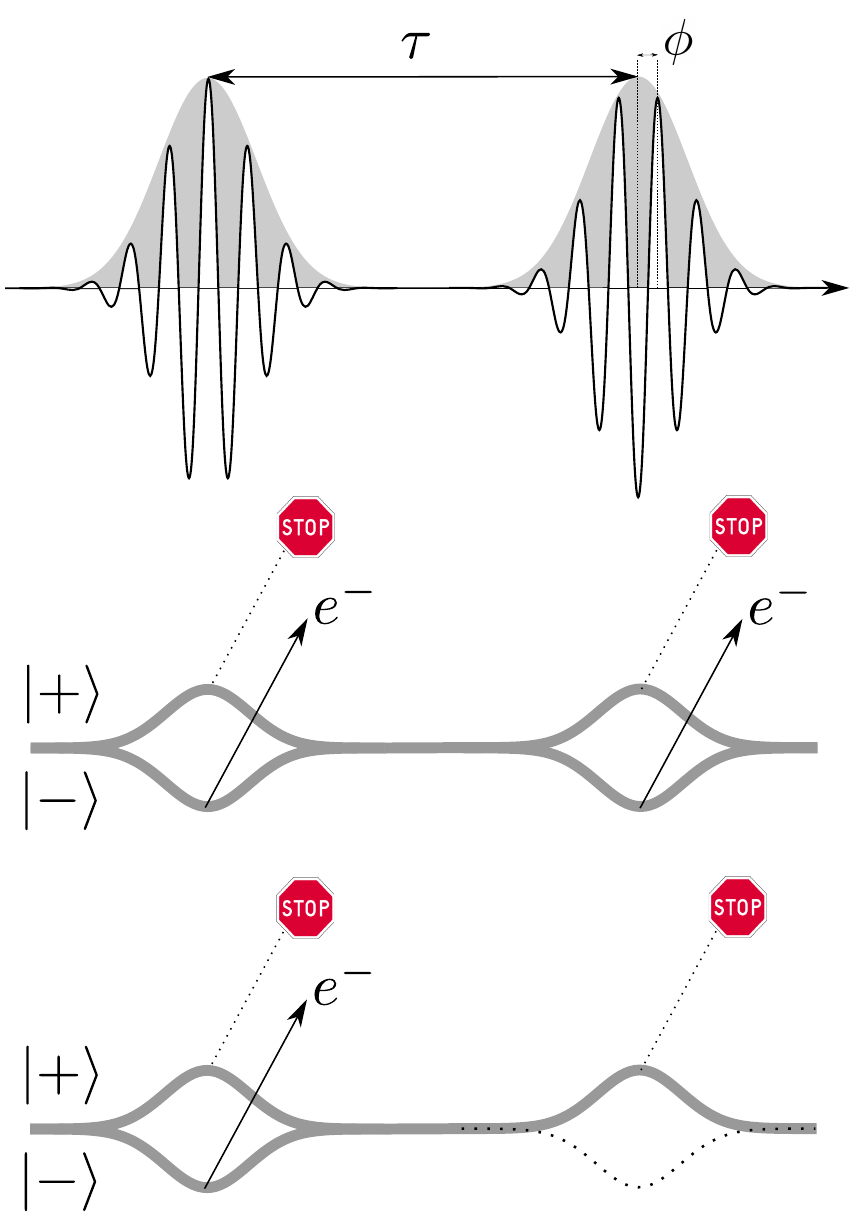}
    \caption{Conceptual picture of the proposed pump--probe scheme with a delay $\tau$. A pump pulse \EdvinUpdate{resonant with an atomic transition} prepares the atom in a superposition of dressed states, $\ket{+}$ and $\ket{-}$ (left side), \EdvinUpdate{which are eigenstates of the combined atom-field system. When the field is present, the dressed states separate into upper, $\ket{+}$, and lower, $\ket{-}$, energy branches. Due to interfering ionization processes,} the atom can photoionize from the $\ket{-}$ state, but not from the $\ket{+}$ state (marked by STOP signs), as shown in the middle cartoon.  By controlling the CEP phase of the probe pulse $\phi$,  the final population can be targeted to either dressed state (right side). If mostly the $\ket{+}$ state is populated during the second pulse then the total amount of ionization will be suppressed, as shown in the lowest part of the cartoon.}
    \label{fig:scheme}
\end{figure}

In this paper, we show how QS in two-photon resonant ionization (1+1), a phenomenon first predicted by Beers and Armstrong in 1975 \cite{Beers1975}, can be \EdvinUpdate{used to interpret} a DS phenomenon using a sequence of two pulses, as schematically shown in Fig.~\ref{fig:scheme}.  
Our general ideas are inspired by prior works on alkali atoms by Wollenhaupt {\it et al.} from 2003 \cite{Wollenhaupt2003}, where two optical laser pulses were used to create complex photoelectron patterns by applying coherent control with two dressed states. We study, instead, helium atoms in the extreme ultraviolet (XUV) regime, where the conditions required for QS are feasible \cite{olofsson_photoelectron_2023}, thanks to the development of seeded short-wavelength free-electron lasers (FELs) \cite{allaria_highly_2012}. Recently, nonlinear interference phenomena \cite{Nandi2022} and quantum entanglement \cite{nandi_generation_2024} have been detected for XUV-FEL pulses, and coherent control of ionization probabilities has been reported for chirped XUV-FEL pulses \cite{richter_strong-field_2024}, opening the field of short-wavelength strong-coupling physics on ultrashort timescales. The latter phenomenon arises due to a DS process predicted by Saalmann {\it et al.}\ in 2018 \cite{Saalmann2018}, which can be interpreted using Beers and Armstrong's QS regime, because a stabilized dressed state can be selected using chirped pulses by rapid adiabatic passage. 
We predict that coherent control of atomic stabilization can alternatively be achieved by using a sequence of two unchirped, phase-locked FEL pulses, provided that their intensities, pulse durations, polarizations, and photon energies are appropriately chosen, in accordance with the QS results reported in Ref.~\cite{olofsson_photoelectron_2023}. 

Our article is organized as follows: an overview of atomic stabilization phenomena is given in Sec.~\ref{sec:overview}, followed by an outline of our theory, with ionization rates extracted from quasienergy calculations in Sec.~\ref{sec:theory}, our results for time-dependent population dynamics and coherent-control of stabilization against photoionization in Sec.~\ref{sec:results}, and our conclusions in Sec.~\ref{sec:conclusions}. \EdvinUpdate{There are also two appendices that discuss gauge invariance for restricted basis set calculations, App.~\ref{app:gau_inv}, and how dipole moments differ between configuration interaction singles and single active electron models, App.~\ref{app:sqrt}.} 
Atomic units are used unless otherwise stated: $\hbar=e=4\pi\epsilon_0=m_e=1$.

\section{Overview of stabilization phenomena}
\label{sec:overview}
In the following section, we present a summary of different stabilization phenomena, \EdvinUpdate{both QS and DS,} that have been studied during the last decades.
While the more general notion of {\it bound states in the continuum} goes back to early work on quantum mechanics by Wigner and von Neumann from 1929 \cite{wightman_uber_1993}, the subject has expanded to atoms in magnetic fields by Friedrich and Wintgen in 1984 \cite{friedrich_physical_1985}, to optics with resonators and photonic structures, as reviewed by Koshelev {\it et al.}\ \cite{koshelev_bound_2023}, and to scattering theory, as reviewed by Domcke \cite{domcke_theory_1991}. 
The fact that atoms can stabilize in strong fields under vastly different conditions has also attracted much independent interest in the strong-field community.
It is not possible here to account for all this development, but we aim to stress, by giving a few examples, that strong-field stabilization of atoms is a broad subject, beyond the superintense regime discussed by Gavrila \cite{gavrila_atomic_2002}. After our brief historical account, we then proceed with the main results and discussions about the stabilization regime of interest in this work.


\subsection{High intensity and low frequency}
The first theoretical line was initiated in 1964 by Keldysh, who proposed a unified theory of non-resonant photoionization ranging from the tunneling regime to the multiphoton regime \cite{Keldysh:1965ojf}. In this approach, the Keldysh parameter is introduced as $\gamma_\text{Keldysh}=\sqrt{I_p/2U_p}$, where $I_p$ is the binding potential of the atom and $U_p=E_0^2/4\omega_0^2$ is the ponderomotive energy of the electron in the laser field. Photoionization described by SFA is known as Keldysh--Faisal--Reiss (KFR) theory \cite{Keldysh:1965ojf,faisal_multiple_1973,reiss_effect_1980}, with free electrons driven by the instantaneous laser field without 
the atomic potential. The SFA is a widely successful theory that has provided key insights for a plethora of strong-field phenomena, including above-threshold ionization (ATI) \cite{agostini_free-free_1979} and high-order harmonic generation (HHG) \cite{ferray_multiple-harmonic_1988}, as reviewed by Amini {\it et al.}\ \cite{amini_symphony_2019}. A new neutral exit channel for tunneling ionization was discovered in 2008 by Nubbemeyer {\it et al.}\ \cite{nubbemeyer_strong-field_2008}. The effect relies on recapture of the electron by Rydberg states after tunneling ionization, a process beyond SFA \cite{popruzhenko_quantum_2018,hu_quantum_2019,olofsson_frustrated_2021}. The physics of such {\it frustrated tunneling} (FT) is efficiently understood by sampling quasi-classical electron trajectories that tunnel from the atom and then are driven in the combined non-conservative laser-atom potential that leads to trapping in Kepler-like orbits \cite{shvetsov-shilovski_capture_2009}. There is strong dependence on the exact pulse shape and duration, which implies that FT can be considered a DS process. The transition from strong-field to multiphoton excitation has been studied using the channel-closing mechanism in the time and frequency domains for neon and argon atoms \cite{zimmermann_unified_2017}. Typical values for laser parameters in low-frequency high-intensity experiments are $>3\times 10^{14}$ W/cm$^2$ at 800 nm, and given the binding energy $13.6$ eV of a hydrogen atom, this then corresponds to a Keldysh parameter in the tunneling regime: $\gamma_\text{Keldysh}< 0.6< 1$.  


\subsection{High intensity and resonant frequency}
The second theoretical line of strong-field ionization was initiated in 1975 by Beers, Armstrong, and Feneuille to describe resonant photoionization in the strong-coupling (SC) regime \cite{Beers1975,armstrong_resonant_1975}. In the SC regime, light-matter interactions can be understood within the RWA using dressed states \cite{cohen-tannoudji_absorption_1969,CT1998}. 
Beers and Armstrong \cite{Beers1975} and Holt {\it et al.}\ \cite{Holt1983} \EdvinUpdate{used effective Hamiltonian theory} to find that the intensity dependence of the ionization probability can be characterized in terms of three system parameters: the ionization rates of the essential atomic state, $\ket{a}$ and $\ket{b}$, denoted $\Gamma_a$ and $\Gamma_b$, respectively, and their mutual coupling coefficient, $q$. We note that $q$ accounts for both the direct transition between the essential states and higher-order virtual transitions via non-essential states (including the continuum). It was shown that atoms can photoionize in different ways, {\it e.g.}, the atom may ionize primarily from the ground or excited atomic state. It was also proposed that quantum interference between these ionization pathways can play a role in the dynamics. In particular, it was predicted that there exist special parameters for which the ionization probability of the atom is bounded by some value $C$ such that $P_{\textrm{ion}} \leq C < 1$ when the atom is subjected to a monochromatic field. For the resonant case one requires that $\Gamma_a\approx\Gamma_b$, which means that the atom ionizes with similar rates from both atomic states. 
One limitation of these early works  was, however, that they only considered ionization to a single continuum \cite{Beers1975,Holt1983}. 
 In 2023, the theory of Beers and Armstrong was generalized to multiple continua by Olofsson and Dahlstr\"om and applied to the experimentally relevant helium atom at the XUV transition: 1s$^2$-1s2p \cite{olofsson_photoelectron_2023}. It was found that {\it dressed-atom stabilization} was reached at an intensity of $\sim 10^{14}\,$W/cm$^2$. In hindsight, the fact that atoms can photoionize at the same rate from both essential states in strong coupling was identified experimentally already in 2022 by Nandi {\it et al.} for the $1s^2-1s4p$ transition at $2\times 10^{13}$ W/cm$^2$ using seeded XUV-FEL pulses with linear polarization \cite{Nandi2022}. More recently, experiments have been performed on the $1s^2-1s2p$ transition using circular polarized XUV-FEL pulses at intensities in the range of $10^{14}$ W/cm$^2$ by Richter {\it et al.} \cite{richter_strong-field_2024}.  
Related stabilization effects, due to destructive interference of two leaky sources, have previously been found for atoms in strong magnetic fields \cite{friedrich_physical_1985} and for light propagation in photonic structures \cite{koshelev_bound_2023}. 
Dressed-atom stabilization physics corresponds to Keldysh parameters in the multiphoton regime, $\gamma_\text{Keldysh}\approx 10>1$.  


\subsection{High intensity and high frequency}
At more extreme conditions, both time-independent Floquet approaches and time-dependent numerical propagation approaches have been adopted to study atoms in superintense high-frequency laser fields. In this case,  one photon is enough for photoionization. It was shown, e.g., by Eberly and Kulander in 1993 \cite{eberly_atomic_1993}, that atoms could undergo counterintuitive dynamics and become more stable as an ionizing pulse gets more intense. The effect is now referred to as {\it superintense atomic stabilization},  \cite{pont_stabilization_1990,gavrila_atomic_2002}.
The original idea to stabilize (hydrogen) atoms in their ground state has remained out of reach, due to simultaneous requirements of high intensity and photon energy, such as $3\times 10^{16}\,$W/cm$^2$ and $\omega_0=27.2\,$eV (one in atomic units). This leads to a Keldysh parameter close to unity for the binding energy of the hydrogen atom, $\gamma_\text{Keldysh}=\sqrt{2}$.


\subsection{Stabilization of autoionizing resonances}\label{sec:auto_intro}

Finally, we mention that autoionizing states in atoms, subjected to resonant strong laser fields, can lead to stabilization of one of the polaritonic branches (i.e., dressed states), as first predicted by Lambropoulos in 1981  \cite{lambropoulos_autoionizing_1981}. Recently, such effects have been studied in experiments 
and theory developed further \cite{kylstra_double_1998,harkema_autoionizing_2021,yanez-pagans_multipolariton_2022,cariker_autoionizing_2024}. The underlying physics that leads to stabilization of autoionizing polaritons is closely related to dressed-atom stabilization, since both phenomena arise due to the interference of different ionization processes within the RWA. The physical origin of autoionization can be a static configuration interaction \cite{lambropoulos_autoionizing_1981}, induced dynamically by the same field that strongly couples the atoms \cite{Beers1975}, or by an additional laser field that couples to a continuum \cite{dai_laser-induced_1987}. If the laser envelope shape can be coherently manipulated, the latter case opens up for independent and time-dependent control of autoionization processes and DS phenomena. 


\begin{figure*}
    \centering
    \includegraphics[width=0.8\linewidth]{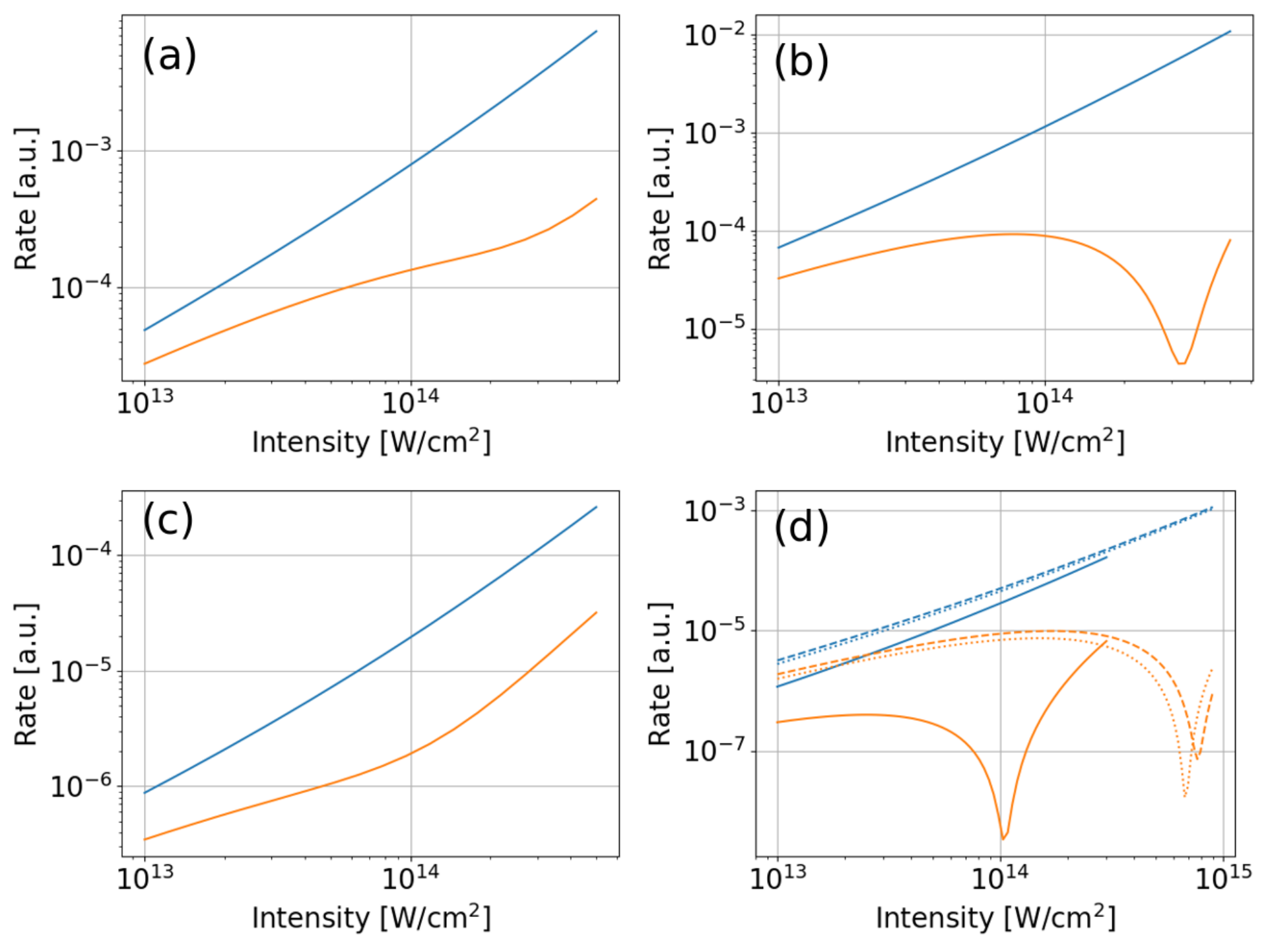}
    \caption{Dressed-state ionization rates extracted from the imaginary part of the corresponding quasienergies. In all cases, the blue curves correspond to the $\ket{-}$ dressed state, and the orange ones to the $\ket{+}$ dressed state.  Linear polarization is used for (a) and (c), and circular polarization for (b) and (d). The top row shows results for hydrogen (with $\omega = 0.375$ a.u.), and the bottom row for helium. The calculations for He were performed with an $\omega$ corresponding to the field-free resonance for each model of He, CIS ($\omega = 0.79721$ a.u., solid lines), $V^{\textrm{He}}_1$ ($\omega = 0.78118$ a.u., dotted lines), and $V^{\textrm{He}}_2$ ($\omega = 0.77679$ a.u., dashed lines).}
    \label{fig:rates}
\end{figure*}
\section{Theory}\label{sec:theory}
In the following subsections, we provide a brief account of the theory applied in this work.

\subsection{Form of the laser pulses}\label{sec:pulses}
\EdvinUpdate{
In this work, the total laser vector potential is given by
\begin{equation}
    A(t) = A_{\textrm{pump}}(t) + A_{\textrm{probe}}(t),
\end{equation}
and the electric field is $E(t)=-\dot A(t)$.
The pump and probe pulses are defined by
\begin{equation}
    A_{\textrm{i}}(t) = A_0f(t)[A_x(t),A_y(t),A_z(t)],
\end{equation}
where $A_0$ is the amplitude of the vector potential and $f(t)$ is the pulse envelope
\begin{equation}\label{eq:env}
    f(t) = \begin{cases}\cos^2\left[\frac{\pi(t-t_0)}{\tau}\right], & |t-t_0|<\frac{\tau}{2}\\
            0, &|t-t_0|\geq\frac{\tau}{2}.
    \end{cases}
\end{equation}
In Eq.~\eqref{eq:env}, $t_0$ and $\tau$ correspond to the pulse delay and pulse duration (foot-to-foot), respectively.
In the case of linear polarization, the field components are chosen as
\begin{align}
\begin{split}
      &A_x(t) = 0,\\
      &A_y(t) = 0, \\
      &A_z(t) = \sin\left[\omega (t-t_0) -\varphi  \right],
\end{split}
\end{align}
where $\varphi$ is the carrier--envelope phase (CEP).
For circular polarization the components are given by
\begin{align}
\begin{split}
      &A_x(t) = \frac{1}{\sqrt{2}}\sin\left[\omega (t-t_0) -\varphi  \right],\\
      &A_y(t) = -\frac{1}{\sqrt{2}}\cos\left[\omega (t-t_0) -\varphi\right], \\
      &A_z(t) = 0.
\end{split}
\end{align}
The probe pulse is delayed by a fixed amount $t_0>\tau$ with a variable CEP $\varphi\in[0,2\pi]$, while the pump pulse is fixed to give a zero reference at $t_0^\mathrm{pump}=0$ with $\varphi^\mathrm{pump}=0$. In this configuration, we define the {\it field phase} of the probe field
\begin{equation}
\phi=\omega t_0+\varphi.
\label{eq:fieldphase}
\end{equation}
}

\subsection{Two-level system dynamics}
\EdvinUpdate{We will now provide a brief discussion of the dynamics of a two-level system that is relevant for our proposed scheme. For simplicity, we will work with linear polarization, but the ideas also apply to circular polarization. In this case, the electric field of a single pulse takes the form $E(t)=\-\vec E_0(t)\cos(\omega t-\phi)$, where $E_0(t) = \omega A_0f(t)$ and we have assumed that the envelope varies slowly enough for its derivative to be negligible. The atomic states that make up the two-level system are the ground, $\ket{a}$, and the excited state, $\ket{b}$.} 

As is well known, the dynamics of two-level systems can be interpreted in terms of a {\it torque vector}, $\vec{\Omega}$, that rotates the {\it pseudospin}, $\vec{s}$, on the Bloch sphere according to \cite{feynman_geometrical_1957,
allen_optical_1975}
\begin{equation}
\frac{d}{dt}\vec{s}(t)=\vec{\Omega}(t)\times \vec{s}(t).
\label{eq-torque}
\end{equation}
In the rotating frame, and within the RWA, the dynamics of the two-level atom can be alternatively expressed on a Hamiltonian form using Pauli matrices, $\vec{\sigma}=(\sigma_x,\sigma_y,\sigma_z)$, as 
\begin{equation}\label{eq:Pauli}
i\frac{d}{dt}\ket{\psi}= \frac{1}{2}\vec{\Omega}\cdot \vec{\sigma}\ket{\psi}
\end{equation}
where the torque vector can be written as 
$\vec\Omega=(\EdvinUpdate{-}\Omega\cos\phi,\EdvinUpdate{-}\Omega\sin\phi,\Delta)$ 
in Cartesian coordinates \EdvinUpdate{and our choice of field parametrization}. 
More precisely, the Rabi frequency is calculated in the electric dipole form, \EdvinUpdate{$\Omega(t)=| z_{ba} E_0(t)|$, where $z_{ba} = \bra{b}z\ket{z}$ is the dipole matrix element between the ground and excited states}, and the detuning of the field is taken relative to the atomic transition as $\Delta=\omega-\omega_0$. If we assume that $\Delta=0$, then the torque vector resides in the $xy$-plane and the pseudospin revolves on a great circle from the ground state to the excited state. The amount of revolution is determined by the area theorem, given by the pulse area, $\theta=\int dt \Omega(t)$.  
\EdvinUpdate{The eigenvectors of the Hamiltonian on the right-hand side of Eq.~\eqref{eq:Pauli} have the form
\begin{equation}
\ket{\pm}=\frac{1}{\sqrt{2}}\left(\ket{a}\EdvinUpdate{\mp} e^{i\phi}\ket{b}\right),
\label{eq:dressed}
\end{equation}
and are known as dressed states. Their eigenvalues are given by $\pm\Omega/2$.}

 Assuming that the pump pulse is a $\theta_\mathrm{pump}=\pi/2$ pulse with $\phi_\mathrm{pump}=0$, \EdvinUpdate{the pseudospin after the pump pulse will lie in the $xy$-plane.} The field phase of the probe pulse can then be chosen to make its corresponding torque vector parallel with the pumped pseudospin by $\phi_\mathrm{probe}=\pm\pi/2$. The probe pulse will then not drive any further dynamics on the Bloch sphere, as follows from the cross product in Eq.~(\ref{eq-torque}). There are two such stationary cases, corresponding to parallel and anti-parallel orientation of torque and pseudospin. \EdvinUpdate{This corresponds to the situation where the state after the pump pulse, $\ket{\Psi_{\mathrm{pump}}}$, is one of the dressed states of the probe pulse. In general, $\ket{\Psi_\mathrm{pump}}$ will be a superposition of $\ket{\pm}$, and the relative population can be controlled by changing $\phi_\mathrm{probe}$.}


\subsection{Extracting ionization rates from quasienergies}

The main point of this work is to study how ionization can be controlled by utilizing the different ionization rates of dressed states, beyond the resonant two-level model given in Eq~(\ref{eq:dressed}). While our main ideas stem from the effective Hamiltonian presented in Ref.~\cite{olofsson_photoelectron_2023}, which was computed using perturbation theory, we here compute the dressed-state ionization rates by directly finding the corresponding eigenvalues of the Floquet Hamiltonian. 

Problems in quantum mechanics with time-dependent interactions can be studied using techniques for time-independent Hamiltonians by considering an extended Hilbert where the time coordinate is treated on equal footing to the spatial ones \cite{howland_stationary_1974, yao_stationary_1994}. 
For interactions that are periodic in time, this procedure leads to Floquet theory,  where the atomic Hilbert space is augmented by the space of functions with the same period as the interaction \cite{maquet_stark_1983,chu_beyond_2004}. In the appropriate limit, Floquet theory can be connected to quantum optics, as first shown by Shirley in 1965 \cite{shirleySolutionSchrodingerEquation1965}. 
If the light is circularly polarized, the problem can equivalently be formulated in a co-rotating reference frame \cite{chu_quasienergy_1978, tip_atoms_1983,kjellsson_lindblom_atomic_2021,dubois_energy_2024}, which leads to a correspondence between the Floquet harmonic index and the magnetic quantum number of a given atomic state.

In the Floquet theory approach, the solution to the time-dependent problem can be constructed from the solutions to the eigenvalue problem of the so-called Floquet Hamiltonian \cite{chu_beyond_2004},
\begin{equation}
    H_F\ket{\psi} = \lambda\ket{\psi}. 
\end{equation}
where $\lambda$ are {\it quasienergies} and $\ket{\psi}$ are the corresponding dressed states. 
The Floquet Hamiltonian in length gauge for linear polarization along the $z$-axis reads 
\begin{equation}
    H_F = H_0\otimes \mathbb{1} + \omega\mathbb{1}\otimes D + \frac{E_0}{2}z\otimes F, 
\end{equation}
\EdvinUpdate{where $H_0$ is the field-free atomic Hamiltonian, $\mathbb{1}$ is the identiy matrix, $D =  \mathrm{diag}(-n,\dots,0,\dots,m)$, $z$ is the $z$-component of the atomic dipole operator, $F$ is a matrix with ones on the sub- and superdiagonals, and $\otimes$ denotes the tensor product of the atomic Hilbert space with the Floquet basis.}
For circular polarization in the $xy$-plane, the co-rotating frame formulation leads to the following Hamiltonian
\begin{equation}
    H_F = H_0 -\omega L_z + \frac{E_0}{\sqrt{2}}x,
\end{equation}
where $L_z$ is the $z$-component of the orbital angular momentum operator, and $x$ is the atomic dipole operator along the $x$-axis.

If we apply an exterior complex scaling (ECS) transformation \cite{Simon1979},the quasienergies will be complex,
\begin{equation}
    \lambda = E - i\frac{\Gamma}{2}.
\end{equation}
For isolated eigenvalues, $\Gamma$ can be interpreted as the decay rate of a resonance, analogous to the application of ECS to scattering problems \cite{Simon1979, Moiseyev1998, graffi_exterior_1983, howland_complex_1983, graffi_resonances_1985}. This means that $\Gamma$ represents the photoionization rate of the system, if it was originally prepared in the corresponding dressed state. 

In this work, the atomic states are constructed using the configuration interaction singles (CIS) approximation \cite{foresmanSystematicMolecularOrbital1992, dreuwSingleReferenceInitioMethods2005}, with a B-spline representation for the radial component of the atomic orbitals. The resulting matrix eigenvalue problem is then solved for a few quasienergies close to a specified target.
%


\subsection{Stabilization of dressed states}
 The bound state dynamics of the system we are considering can be well understood using a non-Hermitian two-level effective Hamiltonian that incorporates the decay into the continuum. 
 The corresponding dressed states will be labeled $\ket{\pm}$ with eigenvalues $\lambda_\pm\in \mathbb{C}$. 
 As mentioned in Sec.~\ref{sec:auto_intro}, a ground state resonantly coupled to an autoionizing state, or interfering autoionizing resonances, admits a similar description, since they can also be mapped onto such an effective Hamiltonian.
 Therefore, there exist rich analogies in their respective dynamics \cite{kylstra_double_1998,litvinenko_multi-photon_2021}. For certain choices of parameters, one of the eigenvalues of the effective Hamiltonian has (nearly) vanishing imaginary part, and hence the corresponding dressed state is (nearly) stabilized against ionization \cite{Beers1975,lambropoulos_autoionizing_1981,kyrola_n_1986, kylstra_double_1998}. In the case of autoionizing resonances, the structure in the continuum is induced by configuration interaction \cite{fano_effects_1961}, while in the system studied in this paper, it is induced by a multiphoton transition \cite{Beers1975,armstrong_resonant_1975,litvinenko_multi-photon_2021}. As will be shown in Sec.~\ref{sec:results}, the coupling between the resonant states is typically stronger than the decay into the continuum, which means we are in the strong-coupling regime \cite{limonov_fano_2017}.

As explained in Ref.~\cite{olofsson_photoelectron_2023} the extent to which one of the dressed states can be stabilized depends on the polarization of the light, since this determines the number of partial waves that are accessible for the photoelectron. Physically, the reduction in ionization rate will be smaller if the interference is not destructive for all partial waves simultaneously. For a given initial $s$-orbital, such as from the ground states of hydrogen $1s$ or helium $1s^2$ ${^1S_0}$, the dipole selection rules for circular polarization allow only $d$-wave photoelectrons in two-photon ionization, while both $s$- and $d$-wave photoelectrons are allowed for linear polarization. This explains the differences between linear and circular polarization for the dressed-state ionization curves in Fig. \ref{fig:rates}, where a local minimum in $\Gamma_+$ at a non-zero intensity is present for circular polarization. In contrast, $\Gamma_+$ has no local minimum for linear polarization, but it is suppressed relative to $\Gamma_-$. For hydrogen, the intensity at which the minimum occurs is beyond the perturbative approach of effective Hamiltonian theory, as pointed out in Ref.~\cite{olofsson_photoelectron_2023}, but revealed here by non-perturbative quasienergy calculations. In the case of helium, we present both CIS results, which incorporate the fact that helium consists of two electrons, and single-active electron (SAE) results, based on two different effective one-electron potentials, 
\begin{equation}
    V^{\textrm{He}}_1(r) = -\frac{1}{r}\left(1+e^{-r/r_0} - re^{-r/r_1}\right),
\end{equation}
with $r_0 = 1.07147$ a.u.\ and $r_1 = 0.83072$ a.u.\ \cite{richter_strong-field_2024}, and 
\begin{equation}
    V^{\textrm{He}}_2(r) = -\frac{1}{r}\left[1+\left(1+\beta r/2\right)e^{-\beta r}\right],
\end{equation}
with $\beta = 27/8$ a.u.$^{-1}$\ \cite{hartree_He_pot}.
While the SAE approach qualitatively reproduces the minimum in the ionization rate from the $\ket{+}$ state, its position in intensity is higher by a factor of 6, compared to the two-electron calculation (CIS). 
Surprisingly, this shows that {\it many-body effects} are essential for a quantitative description of the dressed-atom stabilization phenomenon. 
We attribute a part of the difference between the CIS and SAE results to the fact that the ground state orbital is doubly occupied with spin singlet symmetry. In TDCIS, this leads to an extra factor of $\sqrt{2}$ in the equations of motion for the dipole excitation from the ground state to the excited states \cite{greenmanImplementationPRA2010}, \EdvinUpdate{see App.\ \ref{app:sqrt} for a derivation of this fact}, increasing the corresponding ionization rate by a factor of $2$. This is a simple statistical factor due to the fact that either of the two electrons (spin up or down) can be ionized and contribute to the total rate. The same phenomenon also leads to $\sqrt{2}$ times faster Rabi oscillations in TDCIS compared to SAE, but it does not alter the ionization rate from the excited state. \EdvinUpdate{Another potential source of discrepancy between the models could be the shape of the radial orbitals, which will affect the values of the dipole matrix elements.}
We refer the reader to App.\ \ref{app:gau_inv} for a more detailed numerical study of the gauge invariance of the ionization rates in the full calculation and its dependence on state truncation. 
%


\subsection{Time-dependent approaches}

Time-dependent dynamics are computed using both non-relativistic and relativistic {\it ab initio} methods. We utilize methods that have been described elsewhere, such as the Relativistic Time-Dependent Configuration-Interaction Singles (RTDCIS) method \cite{zapata_rtdcis,tahouri_relativistic_2024} and the TDCIS method \cite{greenmanImplementationPRA2010,bertolino_thomasreichekuhn_2022,carlstrom_general_2022} with t-SURFF to obtain photoelectrons \cite{tao_photo-electron_2012}. 
 While our main target of interest is helium, we have performed additional calculations for hydrogen using the Q-prop code \cite{tulsky_qprop_2020} and found consistent results.

\begin{figure}
    \centering
    \includegraphics[width=0.9\linewidth]{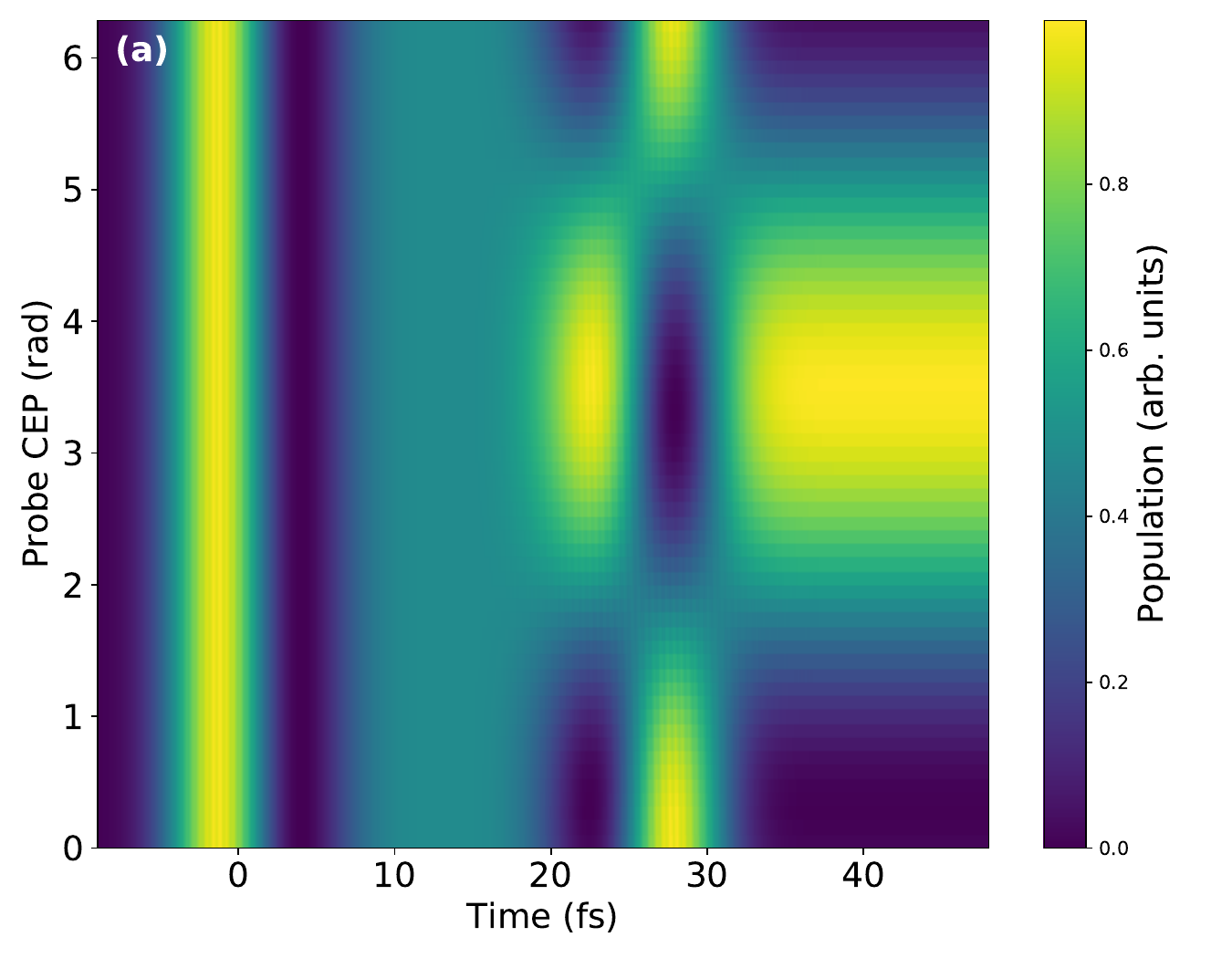}
    \includegraphics[width=0.8\linewidth]{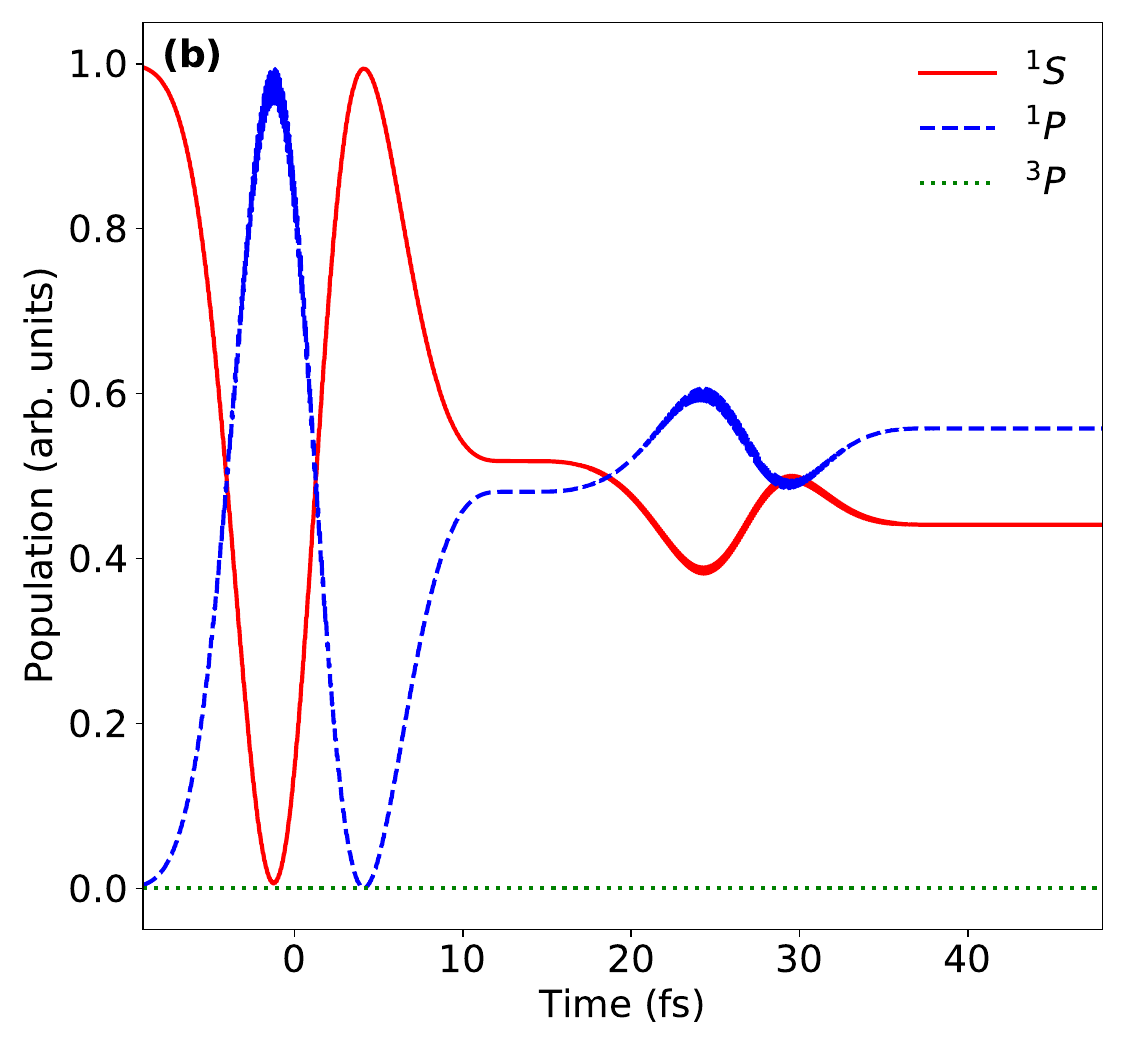}
    \caption{(a) Population of $^1P$ state in helium interacting with a pump-probe scheme of linearly polarized pulses tuned at frequency $\omega=21.7$ eV with the intensity of $I = 5\times 10^{13}$ W/cm$^2$, pulse durations $\tau_1 = \tau_2 = 24.8$ fs, and delay of $t_0 = 26.6$ fs, shown as a function of time and probe CEP. (b) Time evolution of populations of the ground ($^1S$) and excited ($^1P$, $^3P$) states for the fixed probe CEP of $1.26$ rad.}
    \label{fig:rtdcis}
\end{figure}

\section{Results}\label{sec:results}

In general, the atom will be in a superposition of dynamically dressed states, which means that even if one of the states is stabilized, ionization will still take place from the other state. 
If the total amount of ionization is {\it low}, and the field is resonant with the transition, then both dressed states are roughly equally populated. Here, a curious feature is that suppression in ionization from one dressed state is compensated by increased ionization from the other dressed state. This means that no stabilization effect can be detected by monitoring the total amount of ionization from the atom in this case, see further discussion in Ref.~ \cite{olofsson_photoelectron_2023}. 
The effect should still be noticeable as a large asymmetry in the photoelectron AT-doublet, which originates from the different dressed state–continuum coupling strengths \cite{Saalmann2018, Nandi2022, Zhang2022, olofsson_photoelectron_2023}. The asymmetry and ionization probability can be further controlled by steering the relative populations of the dressed states with pulse shaping techniques \cite{Wollenhaupt2006, Saalmann2018}. 

In this work, we control the dressed-state population using two sequential pulses in a pump-probe configuration, following the ideas in Ref.~\cite{Wollenhaupt2003}. The pump pulse prepares the atom in a superposition of two essential states. Control over the {\it field phase} of the probe pulse can then be utilized to selectively populate one of the dressed states during the probe process. If the parameters of the two pulses are chosen such that one of the dressed states is stabilized, then we expect that the field phase of the probe pulse can be used to exert a large degree of control over the ionization dynamics. 
In the following, we present essential state dynamics in Sec.~\ref{sec:results_populations} and  photoelectron dynamics in Sec.~\ref{sec:results-photoelectron}. 

\subsection{Population dynamics}
\label{sec:results_populations}
First, we investigate phase-dependent population dynamics of helium using RTDCIS for the two-pulse case during resonant photoionization (1+1). Linearly polarized pulses, as defined in Sec.~\ref{sec:pulses}, are used. Both pulses are taken to have a pulse area of $\theta\approx 2.5\pi$.
For time-dependent simulations, a complex absorbing potential (CAP) was used instead of ECS to enforce outgoing boundary conditions (since ECS can only be used reliably for time-dependent simulations in the velocity gauge, but there TDCIS suffers from unphysical many-body energy shifts \cite{bertolino_thomasreichekuhn_2022}). 
The form of the CAP used was 
\begin{equation}
V_{\textrm{CAP}}(r) = -i\eta\Theta(r-r_{\textrm{CAP}})(r-r_{\textrm{CAP}})^2,  
\end{equation}
where $\Theta$ is the Heaviside step function, and the parameters were chosen as $\eta = 0.01$ and $r_{\textrm{CAP}} = 110$ a.u. Quasienergy calculations performed with these CAP parameters were able to reproduce the rates shown in Fig.~\ref{fig:rates}(d).

\begin{figure}
    \centering
    \includegraphics[width=1.1\linewidth]{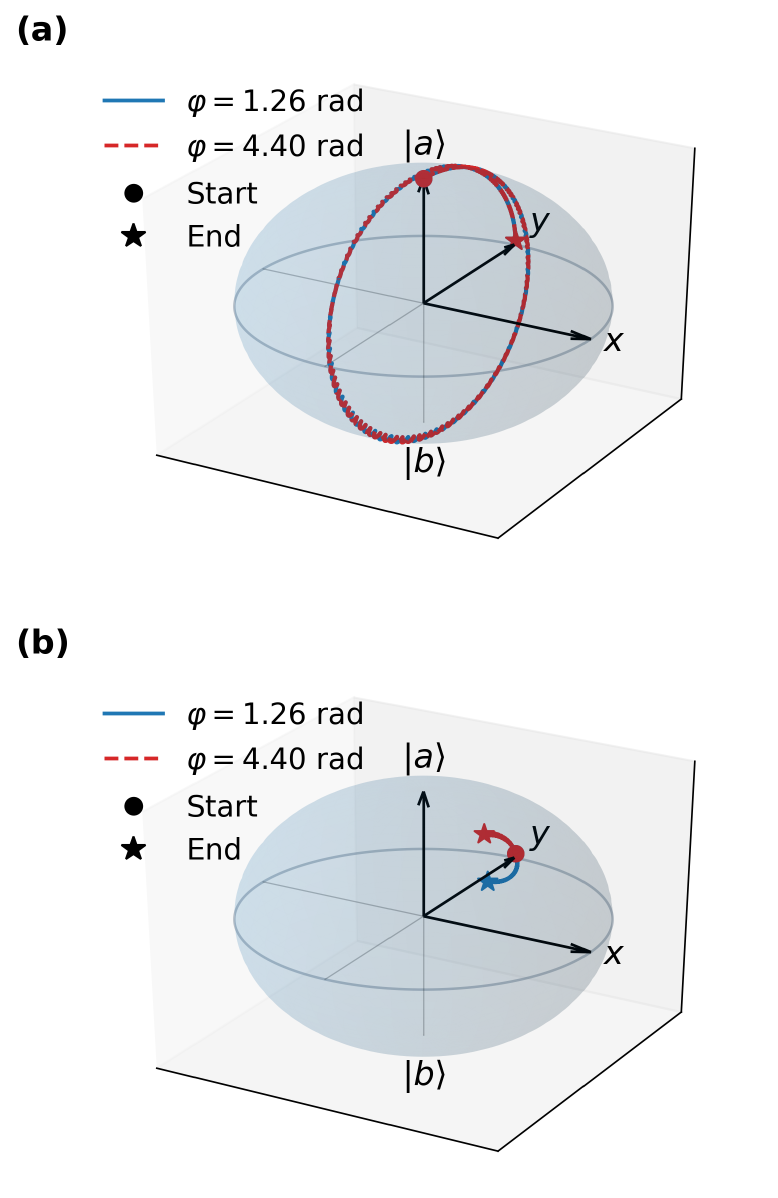}
    \caption{Bloch sphere representation of the state evolution during the pump (a) and probe (b) pulses for two CEP values of the probe pulse: $\varphi = 1.26$ rad (solid) and $\varphi = 4.40$ rad (dotted). The torque vectors of pump and probe pulses are along the $x$ and $y$ directions, respectively (see main text for details).}
    \label{fig:Bloch}
\end{figure}

Here, we are interested in studying how $\varphi$ affects the population dynamics in a helium atom beyond the two-level model at a high intensity. Fig. \ref{fig:rtdcis}(a) shows the population of state $1s2p^1P_1$ as a function of time for different values of probe CEP. The population exhibits a clear CEP dependence, and for $\varphi$ around $1.26$ and $4.40$ rad, the population stays almost unchanged during the probe pulse. These two CEP values correspond to locking the atom on a dressed state with the probe pulse. In Fig. \ref{fig:rtdcis}(b), the populations of $1s2p\,^1P_1$, $1s2p\,^3P_1$, and $1s^2\,^1S_0$ are depicted as a function of time for the probe CEP of $1.26$ rad. It is observed that the populations of ground and excited singlet states undergo modulation with a much reduced amplitude during the second pulse (approximately $10\%$ compared to the Rabi oscillations during first pulse), which implies that a dressed state has been targeted. Using Eq.~(\ref{eq:fieldphase}), the corresponding field phase values are found to be $\sim 0.533\pi$ and $-0.467\pi$ as expected for the ideal dressed states in Eq.~(\ref{eq:dressed}) with $\phi=\pm\pi/2$. The Bloch sphere, illustrated in Fig.~\ref{fig:Bloch}(a), provides an intuitive picture of the population dynamics induced by the pulses. The pump pulse with an area of close to $2.5\pi$ drives the population around the $x$-axis and prepares a nearly equal superposition of ground and excited states. Fig.~\ref {fig:Bloch}(b) shows the subsequent dynamics during the probe pulse. Even though the probe pulse has an area of $2.5\pi$, the populations corresponding to the two CEP values $\varphi=1.26$ and $4.40$~rad (approximately $\pi$ apart) move only slightly because the pump torque is along the $y$-axis (close to the pumped pseudospin directions). Further, it is observed that the induced probe motion is in opposite directions due to the different signs of the torque vector for the two CEP cases. Although deviations from a perfect two-level system arise due to Stark shifts and ionization, the main mechanism of selective dressed-state population is clearly verified by the RTDCIS simulations. 

RTDCIS includes all states with single excitations. We find that the triplet state $1s2p\, ^3P_1$ is not populated, which is expected due to the absence of coupling to the singlet ground state because of symmetry restrictions in the present scenario. Hence, there are no qualitative differences between the relativistic and non-relativistic results for helium, at the level of RTDCIS and TDCIS, and it is not necessary to make a detailed comparison between the two methods in this work. 

\subsection{Photoelectron dynamics}\label{sec:results-photoelectron}
\begin{figure*} 
    \centering
    \includegraphics[width=0.9\textwidth]{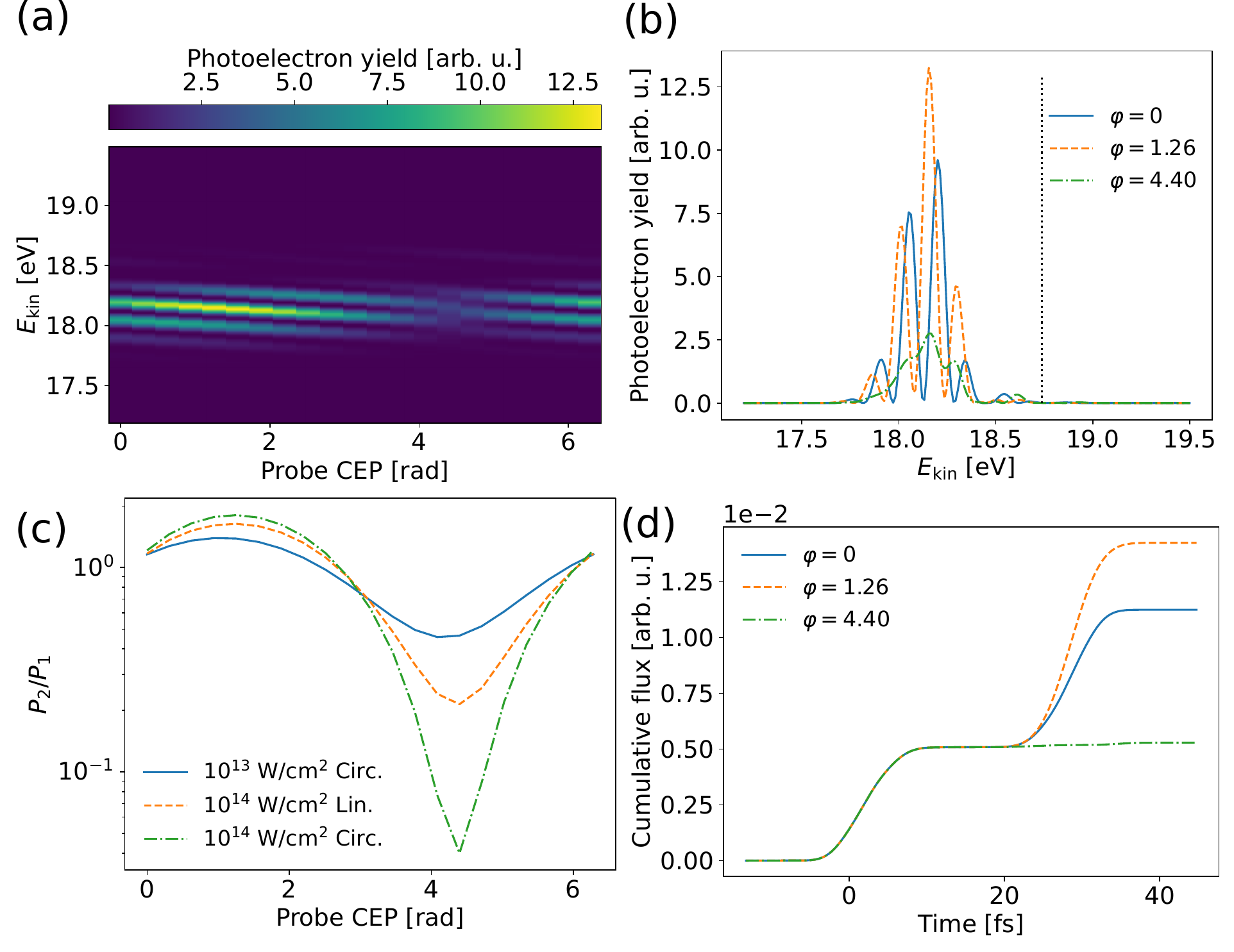}
    \caption{(a) Photoelectron spectra as a function of probe CEP. The pulse parameters are identical to those of Fig.~\ref{fig:rtdcis}, except for the intensity, which in this case is $I=1 \times 10^{14}$ W/cm$^2$. (b) Lineouts from (a) for a few values of the probe CEP. The vertical dotted line indicates the expected location of the upper component ($18.74$ eV) of the AT-doublet based on the location of the lower component ($18.15$ eV) and the Rabi frequency at the peak of the pulse ($0.587$ eV). (c) The ratio of the ionization probability of the probe, $P_2$, and pump, $P_1$, pulses as a function of probe CEP. (d) Cumulative photoelectron flux during the pump-probe sequence, for the same values of the probe CEP as in (b).}
    \label{fig:CEP}
\end{figure*}

In this subsection, we present numerical results for energy-resolved photoionization of helium atoms for the phase-locked two-pulse, resonant two-photon ionization (1+1) process, as shown in 
Figure \ref{fig:CEP}. As will be explained below, our results show that control of the dressed state population by $\varphi$ manifests itself not only in the energy-resolved photoelectron signal, but also in the total amount of ionization. 

The simulations were performed with TDCIS and t-SURFF in the length gauge using both linear and circular polarization. Both pulses have the same duration, but the intensity is increased to meet the stabilization criterion for circular polarization for helium at $10^{14}$\,W/cm$^2$, see  Fig.~\ref{fig:rates}~(d). The resulting pulse areas are approximately $3.5\pi$, i.e., the required $\pi/2$-pulse type, leading to nearly equal populations of the two essential atomic states by the pump pulse.  
Photoelectron spectra were calculated using the Zonte t-SURFF implementation \cite{zonte_repo}. The surface flux was evaluated at a radius of $r=92$ a.u. with a maximum single particle angular momentum of $\ell_{\textrm{max}}=7$. 
%

%

\subsubsection{Photoelectron spectra}
In Fig.~\ref{fig:CEP} (a), the photoelectron energy spectrum is displayed as a function of the CEP of the probe pulse. The first thing one should notice is that, even though the atom undergoes Rabi cycles during the interaction with the pump (and the probe for some CEP), there is only one main photoelectron peak. The usual two peaks of an Autler-Townes doublet, separated by the Rabi frequency $\approx 0.6$ eV, are replaced by a single peak originating from the lower dressed state, $\ket{-}$. This is a {\it signature} of dressed-atom stabilization \cite{olofsson_photoelectron_2023}. However, identifying that the second peak is missing could prove difficult in actual experiments. In a photoelectron spectral measurement, how can one know if the atom is Rabi cycling when the doublet is missing? This is why coherent quantum control is critical for studying the stabilization effect in practice (see also the complementary adiabatic approach of Richter {\it et al.}\ in Ref.~\cite{richter_strong-field_2024}). 
%
%

In our approach, the second pulse introduces additional structures in the photoelectron spectrum. Ramsey-like fringes appear with a spacing that depends on the inverse delay, $t_0>\tau$, between pump and probe pulses, similar to that in Ref.~\cite{Wollenhaupt2003}. This yields an interference signal for photoionization that depends on both the pump and the probe process. The contrast of these fringes gives an indication of the relative ionization probabilities from the two pulses. 
At $\varphi=4.40$, the fringes show less contrast, confirming the reduction in ionization during the probe pulse by targeting the $\ket{+}$ state. Lineouts from Fig.~\ref{fig:CEP} (a) are presented in Fig.~\ref{fig:CEP} (b), showing a clear transition from high contrast ($\varphi=1.26$) to strongly reduced contrast ($\varphi=4.4$). Additionally, the CEP value $\varphi=0$ is found to display nearly {\it maximal} contrast. This can only be attained when the two photoelectron sources are equally strong, which, in our case, implies that the atom must Rabi cycle in the same way driven by both the pump and the probe field. Analysis of this particular CEP value, $\varphi=0$, using Eq.~(\ref{eq:fieldphase}), reveals that it corresponds to a field phase of $\phi = 1.13\pi\approx \pi$. The interpretation is then clear: the probe pulse {\it rewinds} the atomic dynamics back to the ground state, making the ionization process inversion symmetric with that of the pump pulse. Recently, such {\it time symmetries} have been exploited for the control of quantum entanglement in photoionization and in entanglement transfer from ions to spontaneous photons by Stenquist {\it et al.} in Refs.~ \cite{stenquist_harnessing_2025,stenquist_entanglement_2025}. 

We mention that, prior to our work, the case of circular pulses with opposite helicity between the pump and probe has been shown to transform the Ramsey fringes such that they are reduced (or eliminated) without angular resolution. This is because the photoelectron momentum distribution (PMD) would exhibit spiral patterns in the plane of the polarization of the light \cite{fulton_transformation_2024}. Photoelectron spectra along a given azimuth, however, would still show readily apparent fringes in the case that the second pulse ionizes, a pattern against which stabilized results could be contrasted just like for a co-rotating circularly polarized setup. Also, in addition to the AT doublet (or single peak in our case), a third, central (uncoupled) photoelectron peak can appear in the case of two pulses with opposite helicity, as pointed out by Deng {\it et al.} in 2025 \cite{deng_control_2025}, these effects are not present in our scheme because co-rotating pulses are employed.  

\subsubsection{Ionization ratios}
Figure \ref{fig:CEP} (c) displays the ratio $R_{21} = P_2/P_1$ of the ionization probability generated by the pump $P_1$ and probe $P_2$ pulses, as a function of the CEP of the probe pulse. The ionization probability $P_1$ is computed for an isolated pump pulse, while the ionization from the probe pulse is computed as $P_2=P_\textrm{tot}-P_1$, where $P_\textrm{tot}$ is the total amount of ionization from pump and probe pulses. Different combinations of intensity and polarization are considered. This figure demonstrates features that can be interpreted in terms of the dressed-state ionization rates of Fig.~\ref{fig:rates}. The values of $R_{21}$ span more than one order of magnitude for circular polarization at $10^{14}$ W/cm$^2$. 
As could be expected, we find much less control of the ionization at the lower  intensity of $10^{13}$ W/cm$^2$ (in this case we are not in the stabilization regime, but the pulse length was adjusted to yield a $\pi/2$-type pulse).
The value of $\varphi$ where $R_{21}$ is maximal coincides with where the Ramsey fringes in the lower AT-doublet component are most pronounced, indicating that the $\ket{-}$ state is being populated by the probe pulse. The reverse is true for the minimum; it coincides with where the contrast of the fringes is the lowest, which indicates that the $\ket{+}$ state is populated. At $\varphi=0$ (and $\pi$), it can be observed that the amount of ionization is at an intermediate level, which implies that both dressed states are populated during the probe pulse and that the atom resumes Rabi oscillations. 

Finally, the linearly polarized case in Fig.~\ref{fig:CEP}~(c) with $10^{14}$ W/cm$^2$ has the same pulse area, $\theta\approx 3.5\pi$, as the circular case with the same intensity. This makes it a good test case for the theory of stabilization. As should be expected from Fig.~\ref{fig:rates}~(c) and (d), the linearly polarized case shows a reduced amount of control over ionization, which we attribute to the multiple continua (s- and d-waves) that are coupled from the essential states. The fact that the difference is not larger between the two cases is probably due to our choice of a smooth $\cos^2$-envelope function in Eq.~(\ref{eq:env}), while a stronger stabilization would likely arise with a more rectangular pulse. A systematic study of such envelope effects could be conducted using the super-Gaussian sequence, as was recently done for ultrafast absorption and resonance fluorescence at XUV wavelengths \cite{stenquist_mollow-like_2024}, but such a detailed analysis remains beyond the scope of the present work.

\subsubsection{Ionization over time}
Finally, Fig.~\ref{fig:CEP} (d) shows the time-resolved photoelectron flux generated by the pump-probe sequence for the same values of $\varphi$ included in panel (b). It is observed that  photoionization from the probe pulse can be controlled with the suppression {\it or} enhancement of the total yield. Field phases separated by $\pi$ give maximal enhancement or suppression. This means that the energy resolution of the photoelectron is {\it not} required to observe the proposed non-adiabatic stabilization phenomenon. In practice, this is important because the Ramsey-like fringes are likely to be washed out by macroscopic averaging effects in real experiments.  

In summary, if the probe field phase is chosen such that mainly the stabilized dressed state is populated, then ionization will be suppressed during the duration of the probe pulse. Conversely, if the phase is such that the other dressed state is occupied, ionization will instead be enhanced during the probe pulse. If the phase is chosen such that Rabi cycling will continue (or be reversed) during the probe field, the atom will photoionize roughly as it did during the pump field, as was schematically shown in our introduction by  Fig.~\ref{fig:scheme}. 


\section{Conclusions}\label{sec:conclusions}
Atomic stabilization in strong laser fields is a subject that has attracted much attention due to its counterintuitive nature. In this work, we have provided a scheme, based on two pulses, to study non-adiabatic dressed-atom stabilization using fields in the critical intensity regime with circular polarization. The stabilization effect is clearly observed in our {\it ab-initio} numerical simulations by calculating the ionization probability induced by the probe pulse. 
The electric field phase of the probe pulse is a critical parameter that can be used to control stabilization or induce increased ionization by the probe pulse. 
In the energy domain, a single photoelectron peak is observed, despite the atom undergoing Rabi dynamics, with finer Ramsey-like interference fringes appearing on the photoelectron peak. 
The contrast of such fringes is an alternative measure of stabilization that may be difficult to observe experimentally for macroscopic gases. Our work contributes to the expanding subject of bound states in the continuum, and in particular to the role of multiple continua in strongly driven open quantum systems, which are now feasible to study with short-wavelength FELs.   

\begin{acknowledgments}
We acknowledge Saikat Nandi, Yijie Liao, Jakob Bruhnke, Axel Stenquist, and Ulf Saalmann for useful discussions. JMD acknowledges support from the Olle Engkvist Foundation: 194-0734, the Knut and Alice Wallenberg Foundation: 2019.0154 and 2024.0212, and the Swedish Research Council Grant No. 2024-04247. JMND and ELF acknowledge support from the US Department of Energy (DOE), Office of Science, Basic Energy Sciences (BES), under Award No. DE-SC0021054, and the U.S. National Science Foundation under Grant No.PHY-2208078.
\end{acknowledgments}

\appendix
\section{Gauge invariance \EdvinUpdate{and restricted calculations}}\label{app:gau_inv}
\begin{figure}
    \centering
    \includegraphics[width=1.0\columnwidth]{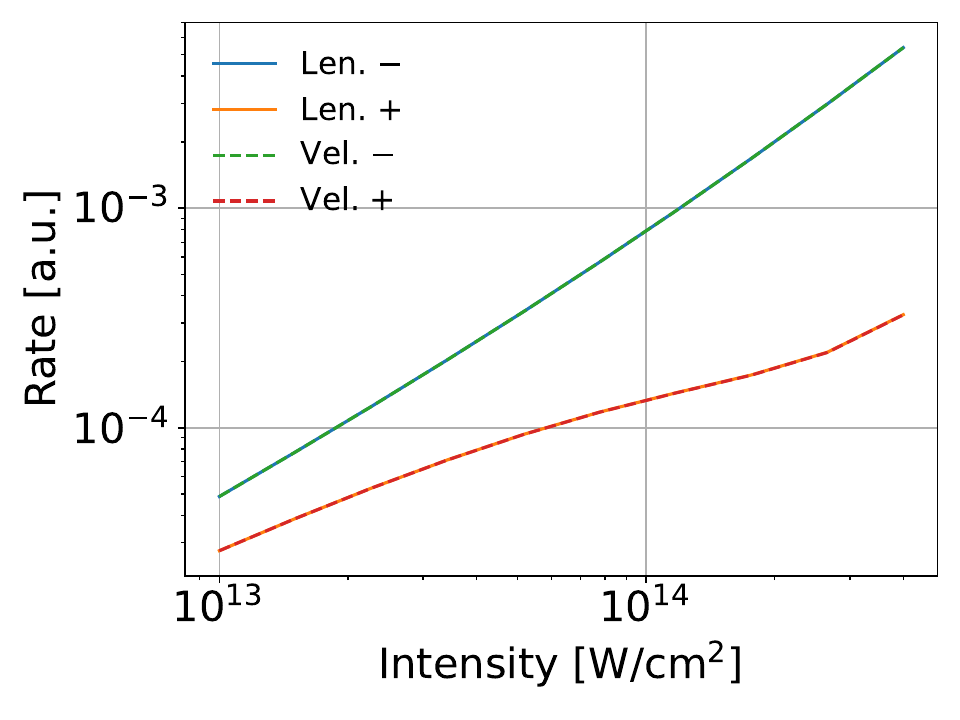}
    \caption{Dressed-state ionization rates for hydrogen as a function of intensity. The solid lines were computed using the length gauge, and the dashed lines were computed using the velocity gauge.}
    \label{fig:gau_inv}
\end{figure}

\begin{figure*}
    \centering
    \includegraphics[width=1\textwidth]{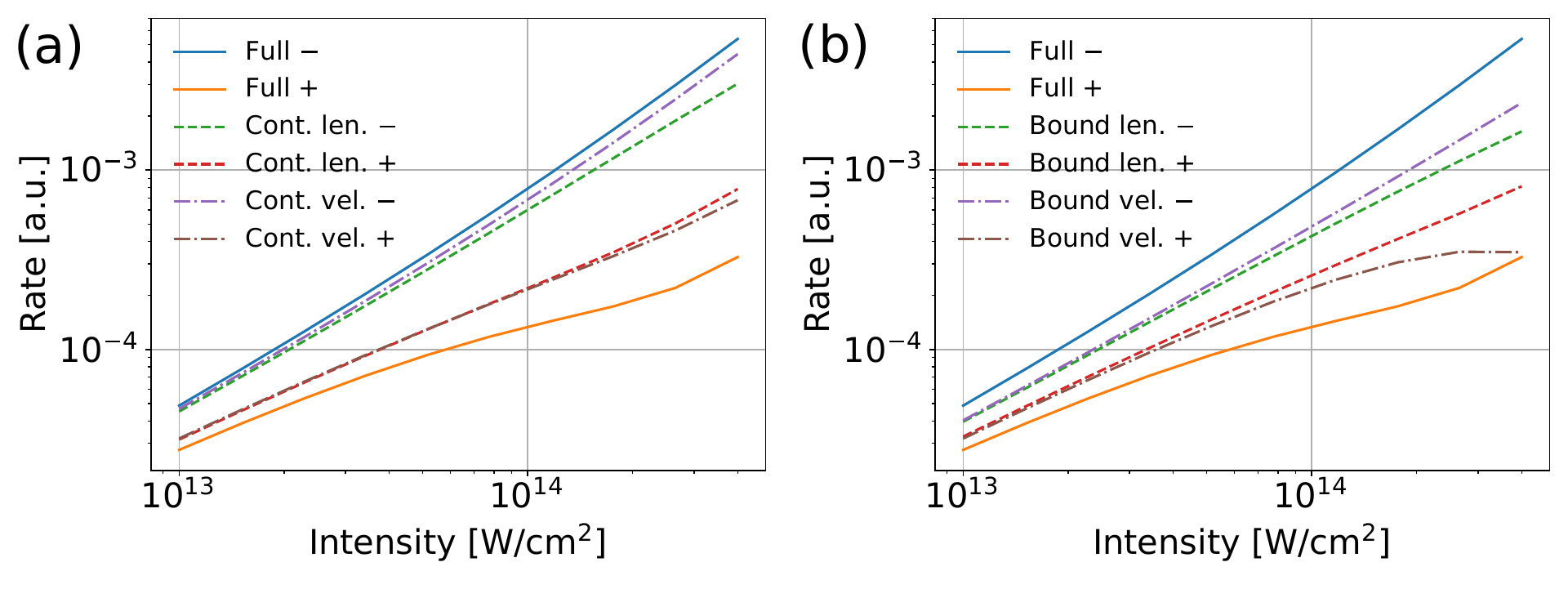}
    \caption{Comparison of ionization rates resulting from full calculations in Hydrogen (solid lines), to calculations where certain states of $p$-symmetry have been removed for length gauge (dashed lines) and velocity gauge (dash-dotted lines). (a)Ionization rates with $p$ bound states other than $2p$ removed. (b) Ionization rates with $p$ continuum states removed.}
    \label{fig:partial_calcs}
\end{figure*}

In Figure \ref{fig:gau_inv} we demonstrate the gauge invariance of the Floquet ionization rates for linear polarization in hydrogen. 
Given that the stabilization mechanism has received renewed recent attention, described as an adiabatic passage to the continuum \cite{Saalmann2018,richter_strong-field_2024}, it may be of interest to examine if the phenomenon can be interpreted as a purely continuum process. In this appendix, we study the effects of truncating the virtual space to explore the role of continuum and bound states separately. However, any such truncation implies an incomplete calculation, with loss of gauge invariance, meaning that the outcome should be interpreted with some care. For this reason we have performed the virtual-space truncations in both length and velocity gauge and compared the results.  
In Fig.~\ref{fig:partial_calcs}, we study the effect on the ionization rates of removing states from the $p$-symmetry of hydrogen. This is done for both the length gauge and the velocity gauge. In panel (a), all bound states of $p$-character except $2p$ have been excluded from the atomic Hilbert space, while in panel (b) all continuum states with $p$-character have been excluded. In both cases, the rates computed with the full atomic Hilbert space are included for comparison. As is evident from both panels, the results of the restricted calculations are not gauge invariant (neither is TDCIS \cite{sato_gauge-invariant_2018,bertolino_thomasreichekuhn_2022}), and none of them can fully reproduce the true ionization rates. The conclusion is that {\it all} virtual states should be included: The stabilization effect is driven by neither purely bound nor purely continuum dynamics in both gauges.   

\EdvinUpdate{
\section{Comparison of dipole moments between CIS and SAE}\label{app:sqrt}
Consider the two orbitals $\phi_{1s}$ and $\phi_{2p}$. Within CIS and the spin-singlet approximation, we can represent the ground-state and a singly excited state by the following states
\begin{align}
    \ket{\Psi_{1s^2}} &= \frac{1}{\sqrt{2}}\ket{\phi_{1s},\phi_{1s}}(\ket{\uparrow\downarrow} - \ket{\uparrow\downarrow}), \\
     \ket{\Psi_{1s2p}} &= \frac{1}{2}(\ket{\phi_{1s},\phi_{2p}} + \ket{\phi_{2p},\phi_{1s}})\left(\ket{\uparrow\downarrow} - \ket{\uparrow\downarrow}\right).
\end{align}
The matrix element for the $z$-component of the dipole moment $z = z_1 + z_2$, where $z_i$ denotes the dipole moment of electron $i$, is given by
\begin{align}
\begin{split}
    \bra{\Psi_{1s2p}}z\ket{\Psi_{1s^2}} &= \bra{\Psi_{1s2p}}z_1\ket{\Psi_{1s^2}} \\
    &+ \bra{\Psi_{1s2p}}z_2\ket{\Psi_{1s^2}}.
\end{split}
\end{align}
Since $z$ is spin-independent, the spin part of the state simply contributes an overall factor of 2 to the matrix element. The individual terms then give
\begin{align}
    \bra{\Psi_{1s2p}}z_1\ket{\Psi_{1s^2}} &= \frac{1}{\sqrt{2}}\bra{\phi_{2p}}z_1\ket{\phi_{1s}},\\
    \bra{\Psi_{1s2p}}z_2\ket{\Psi_{1s^2}} &= \frac{1}{\sqrt{2}}\bra{\phi_{2p}}z_2\ket{\phi_{1s}},
\end{align}
which leads to
\begin{equation}\label{eq:CIS_dip}
    \bra{\Psi_{1s2p}}z\ket{\Psi_{1s^2}} = \sqrt{2}\bra{\phi_{2p}}z\ket{\phi_{1s}},
\end{equation}
where the electron index has been dropped on the RHS. If we instead consider an SAE model where the orbitals are the same as in CIS, then the corresponding matrix element would give
\begin{equation}\label{eq:SAE_dip}
    \bra{\Psi_{2p}}z\ket{\Psi_{1s}} = \bra{\phi_{2p}}z\ket{\phi_{1s}}.
\end{equation}
Comparing Eq.\ \eqref{eq:CIS_dip} and \eqref{eq:SAE_dip}, we can see the factor of $\sqrt{2}$ that is discussed in the main text. If we consider a third orbital, say $\phi_{3d}$, then we find instead
\begin{align}
    &\bra{\Psi_{1s3d}}z\ket{\Psi_{1s2p}} = \bra{\phi_{3d}}z\ket{\phi_{2p}}\\
    &\bra{\Psi_{3d}}z\ket{\Psi_{2p}} = \bra{\phi_{3d}}z\ket{\phi_{2p}},
\end{align}
and there is no difference between the models.
}

\bibliography{bibliography} 
\end{document}